\documentclass{article}

\usepackage{arxiv}

\usepackage[utf8]{inputenc} 
\usepackage[T1]{fontenc}    
\usepackage{hyperref}       
\usepackage{url}            
\usepackage{booktabs}       
\usepackage{amsfonts}       
\usepackage{nicefrac}       
\usepackage{microtype}      
\usepackage{lipsum}		
\usepackage{graphicx}
\usepackage{natbib}
\usepackage{doi}

\usepackage{amsmath}
\usepackage{amssymb}
\usepackage{algorithm}
\usepackage{algpseudocode}
\usepackage{tikz}
\usepackage{graphicx,subfigure}
\usepackage{makecell}
\usetikzlibrary{shapes.arrows}
\usetikzlibrary{calc}
\usepackage{array,multirow}
\DeclareMathOperator*{\minimize}{minimize}

\usepackage{multicol}

\title{Refined Inverse Rigging: A Balanced Approach to High-fidelity Blendshape Animation}

\date{January 2024} 					

\author{ \href{https://orcid.org/0000-0002-5656-9189}{\includegraphics[scale=0.06]{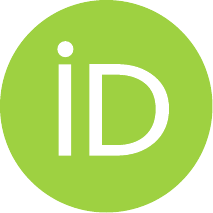}\hspace{1mm}Stevo Racković} \\
	Department of Mathematics\\
	Instituto Superior Técnico\\
	Lisbon, Portugal \\
	\texttt{stevo.rackovic@tecnico.ulisboa.pt} \\
	\And
	\href{https://orcid.org/0000-0003-3071-6627}{\includegraphics[scale=0.06]{orcid.pdf}\hspace{1mm}Cláudia Soares} \\
	Department of Computer Sceince\\
	NOVA School of Science and Technology\\
	Caparica, Portugal \\	
    \And
	\href{https://orcid.org/0000-0003-3497-5589}{\includegraphics[scale=0.06]{orcid.pdf}\hspace{1mm}Dušan Jakovetić} \\
	Department of Mathematics and Informatics\\
	Faculty of Sciences\\
	Novi Sad, Serbia \\	
}

\renewcommand{\shorttitle}{Refined Inverse Rigging}

\hypersetup{
pdftitle={Refined Inverse Rigging: A Balanced Approach to High-fidelity Blendshape Animation},
pdfauthor={Filipa Valdeira, Stevo Racković, Valeria Danalachi, Qiwei Han, Cláudia Soares}
}

\begin{document}
\maketitle

\begin{abstract}
In this paper, we present an advanced approach to solving the inverse rig problem in blendshape animation, using high-quality corrective blendshapes. Our algorithm introduces novel enhancements in three key areas: ensuring high data fidelity in reconstructed meshes, achieving greater sparsity in weight distributions, and facilitating smoother frame-to-frame transitions. While the incorporation of corrective terms is a known practice, our method differentiates itself by employing a unique combination of $l_1$ norm regularization for sparsity and a temporal smoothness constraint through roughness penalty, focusing on the sum of second differences in consecutive frame weights.

A significant innovation in our approach is the temporal decoupling of blendshapes, which permits simultaneous optimization across entire animation sequences. This feature sets our work apart from existing methods and contributes to a more efficient and effective solution. Our algorithm exhibits a marked improvement in maintaining data fidelity and ensuring smooth frame transitions when compared to prior approaches that either lack smoothness regularization or rely solely on linear blendshape models. In addition to superior mesh resemblance and smoothness, our method offers practical benefits, including reduced computational complexity and execution time, achieved through a novel parallelization strategy using clustering methods.

Our results not only advance the state of the art in terms of fidelity, sparsity, and smoothness in inverse rigging but also introduce significant efficiency improvements. The source code will be made available upon acceptance of the paper.
\end{abstract}


\begin{figure}
\centering
    \includegraphics[width=\linewidth]{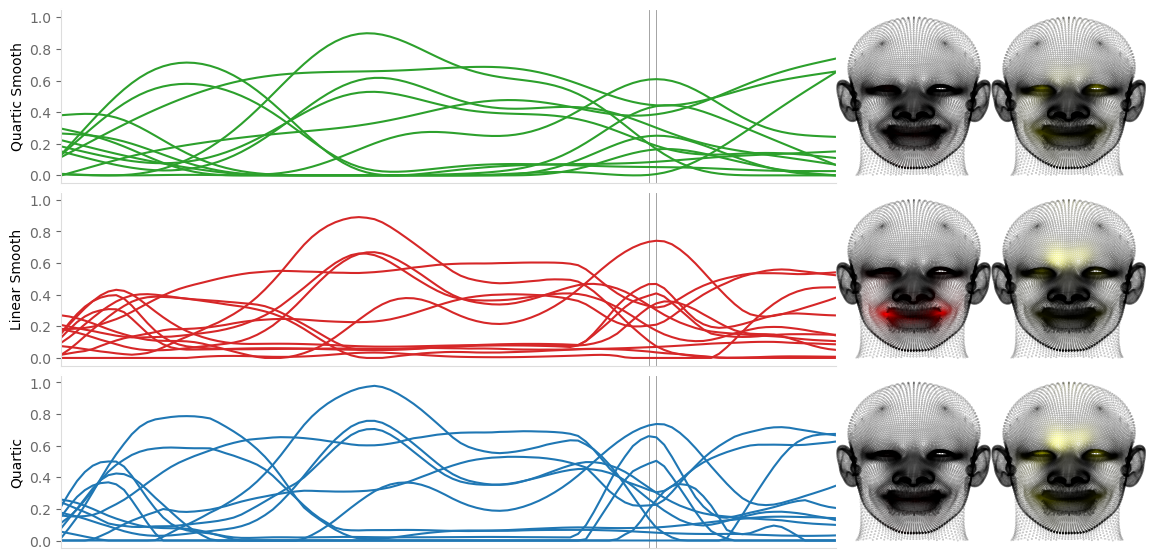}
  \caption{Demonstrating the Efficacy of Temporally Coherent Blendshape Animation. \textbf{First row:} Our \textit{Quartic Smooth} method captures the intricate dynamics of facial expressions by leveraging a sophisticated blendshape rig, ensuring both high-fidelity mesh reconstruction and smooth temporal transitions in animation weights. \textbf{Second row:} The \textit{Linear Smooth} method, as proposed by \cite{seo2011compression}, prioritizes temporal smoothness but simplifies the blendshape model to a linear function, resulting in a trade-off with mesh accuracy. \textbf{Third row:} The \textit{Quartic} approach from \cite{rackovic2023distributed} achieves a high degree of mesh fidelity by utilizing a complex blendshape model but does not account for the smoothness of frame-to-frame transitions, leading to potential discontinuities. Displayed are selected weight trajectories over 100 animation frames, with two consecutive frames magnified to showcase the mesh results. The second column employs red shading to illustrate mesh error, and the third column uses yellow to highlight discrepancies between successive frames, underscoring the balance between accuracy and smoothness in animation sequences.}
  \label{fig:teaser}
\end{figure}

\section{Introduction}

Blendshape animation, a predominant method in animating human faces, manipulates a 3D facial mesh $\textbf{b}_0 \in \mathbb{R}^{3n}$ through interpolation among a predefined set of blendshapes $\textbf{b}_1, \ldots, \textbf{b}_m \in \mathbb{R}^{3n}$, where $n$ denotes the number of mesh vertices \cite{lewis2014practice}. Representing diverse facial configurations, these blendshapes, when linearly combined with weights $\textbf{w} = [w_1, \ldots, w_m]$, enable the creation of a wide range of expressions as
\begin{equation*} \label{eq:LinBlendshapeModel}
    f_L(\textbf{w}; \textbf{B}) = \textbf{b}_0 + \textbf{B}\textbf{w},
\end{equation*}
with $\textbf{B} \in \mathbb{R}^{3n \times m}$ forming a matrix of blendshape vectors. Here, the subscript \textit{L} denotes the linear nature of this blendshape function. Modern advancements incorporate non-linear corrective terms into these models, enhancing realism and flexibility \cite{rackovic2022CD}. For instance, a quartic blendshape function, denoted here as $f_Q$, integrates up to three levels of corrective terms:
\begin{equation}
    f_Q(\textbf{w}) = \textbf{Bw} + \sum_{\{i,j\} \in \mathcal{P}} w_i w_j \textbf{b}^{\{ij\}} + \sum_{\{i,j,k\} \in \mathcal{T}} w_i w_j w_k \textbf{b}^{\{ijk\}} + \sum_{\{i,j,k,l\} \in \mathcal{Q}} w_i w_j w_k w_l \textbf{b}^{\{ijkl\}}.
\end{equation}
Such a detailed correction level is employed in industry-standard solutions like Metahumans\footnote{\url{https://www.unrealengine.com/en-US/eula/mhc}}.

This paper addresses the inverse rig problem: given a target mesh $\widehat{\textbf{b}} \in \mathbb{R}^{3n}$, the goal is to find a set of weights $\textbf{w}$ that accurately approximates the target. Model-based solutions to this problem leverage the structure of rig functions, utilizing optimization techniques rather than relying purely on data \cite{ccetinaslan2016position, rackovic2022majorization}. The state-of-the-art (SOTA) model-based approach, as proposed in \cite{rackovic2022CD}, solves the problem by minimizing the data fidelity of the model while regularizing for sparsity, while constraining the weight values to a feasible set considering that the weights only make sense on an interval $[0,1]$, i.e., 
\begin{equation}\label{eq:holistic_objective}
    \minimize_{\textbf{0} \leq \textbf{w} \leq \textbf{1}} \frac{1}{2} \| f_Q(\textbf{w}) - \widehat{\textbf{b}} \|_2^2 + \alpha \textbf{1}^T\textbf{w},
\end{equation}
employing a coordinate descent method. Prior methods \cite{seol2011artist,lewis2010direct,cetinaslan2020sketching}, while effective, are confined to linear blendshape models and thus are less capable of replicating complex facial dynamics.
However, a critical aspect in animation is the smoothness of frame-to-frame transitions, often overlooked in isolated frame fitting. This temporal dimension has been explored, for instance, by \cite{seo2011compression}, who introduced a smoothness regularization to the optimization objective:
\begin{equation}\label{eq:seo_objective}
    \minimize_{\textbf{0} \leq \textbf{w} \leq \textbf{1}} \| f_L(\textbf{w}) - \widehat{\textbf{b}} \|_2^2 + \alpha \|\textbf{w} \|_2^2 + \beta \| \textbf{w} - \textbf{v} \|_2^2,
\end{equation}
where $\textbf{v} \in \mathbb{R}^m$ represents the weight vector from the previous frame, introducing a temporal continuity constraint. While effective, these methods primarily address linear blendshape models, limiting their capability to capture the more nuanced facial expressions enabled by non-linear models.

In this work, we bridge this gap by proposing a novel objective that harmoniously integrates both the advanced corrective terms of non-linear blendshape functions and the imperative of temporal smoothness. Our formulation extends beyond the scope of \eqref{eq:seo_objective} by concurrently optimizing across all frames. This holistic approach not only ensures the fidelity of each individual frame to the target mesh but also guarantees the smoothness of transitions throughout the animation sequence, by formulating the problem
\begin{equation}\label{eq:our_objective}
    \minimize_{\textbf{0} \leq \textbf{w} \leq \textbf{1}}  \sum_{t=1}^{T} \left( \frac{1}{n}\| f_Q(\textbf{w}^{(t)}) - \widehat{\textbf{b}}^{(t)} \|_2^2 + \frac{\alpha}{m} \|\textbf{w}^{(t)} \|_1 \right)  + \beta \sum_{t=2}^{T-1} \| \textbf{w}^{(t+1)} - 2\textbf{w}^{(t)} + \textbf{w}^{(t-1)} \|_2^2,
\end{equation}
where $T$ is the number of frames in the sequence, and $\beta$ is a regularization parameter that balances data fidelity and smoothness constraints. Notably, our method is adaptable to various levels of corrective terms within the blendshape function, ranging from linear ($f_L$) to higher-order non-linear functions (such as $f_Q$). 

The generalizability and flexibility of our approach enable it to address a wider array of animation challenges, including those with complex facial dynamics. 

This approach, through its comprehensive optimization across the entire animation sequence and its adaptability to various blendshape complexities, presents a significant step forward in achieving more precise facial reconstructions and ensuring smoother motion transitions in blendshape animation.

\section{Related Work}

Facial expressions play a crucial role in human perception and communication, a significance that extends into the realm of 3D animation. This section reviews the evolution and current state of facial animation techniques, particularly focusing on blendshape models, and situates our work within this landscape.

\subsection{Anatomically-Based versus Blendshape Models}
Anatomically-based face models, as discussed in works like \cite{sifakis2005automatic} and \cite{ichim2017phace}, offer high-fidelity deformation and realistic perception. Despite their detailed representation, these models pose challenges in animation control and interpretability. In contrast, blendshape models, a standard in practical face animation due to their simplicity and ease of manipulation, have been extensively explored \cite{Pighin1998SynthesizingRF, choe2001analysis, choe2001performance}. While traditionally sculpted manually, there have been developments towards automation \cite{deng2006animating, buoaziz2013online,moser2021semi,li2010example, li2013realtime, ribera2017facial}. Our work assumes the existence of such a blendshape basis, focusing instead on the inverse rigging process.

\subsection{Inverse Rig Problem and Approaches}
The inverse rig problem, central to our work, involves generating animations by adjusting blendshape weights over time. Two main approaches exist: model-based and data-based. Model-based methods \cite{ccetinaslan2016position, buoaziz2013online, rackovic2022CD} utilize optimization techniques, leveraging the rig structure, as opposed to data-based methods that rely on regression models trained on extensive animated data \cite{Song2011CharacteristicFR, seol2014tuning, holden2015learning, bailey2020fast}. Our approach falls into the former category, focusing on a blendshape-based model-based solution. Previous works in this area typically address least squares problems with additional regularization for stability \cite{ccetinaslan2016position, cetinaslan2020sketching, cetinaslan2020stabilized}, sparsity \cite{buoaziz2013online, neumann2013sparse, rackovic2022majorization}, or temporal smoothness \cite{tena2011interactive, seol2012spacetime}.

\subsection{Direct Manipulation and Face Segmentation}
Related areas include direct manipulation, demanding real-time solutions for adjusting expressions \cite{lewis2010direct, seo2011compression, cetinaslan2020sketching, cetinaslan2020stabilized}, and face segmentation for animation, where the focus varies from creating large mesh segments for inverse rig \cite{joshi2006learning, tena2011interactive, hirose2012creating, reverdy2015optimal, Fratarcangeli2020FastNL, bailey2020fast, rackovic2021clustering} to adding secondary motion effects with smaller segments \cite{zoss2020data, neumann2013sparse, wu2016anatomically, romeo2020data}. While not our primary focus, the concept of face clustering for distributed rig inversion is also explored in our work.

\subsection{Our Contribution}
In this paper, we contribute to the field of blendshape animation by introducing an integrated problem formulation for high-quality rig inversion, combining 1.) complex corrective blendshape terms that enhance the fidelity of mesh reconstructions, drawing on the methodologies established in \cite{rackovic2022majorization, rackovic2023distributed}. We implement 2.) sparsity regularization to achieve a lower cardinality in the weight vectors, which aids in simplifying post-animation adjustments, as discussed in prior works \cite{seol2011artist, rackovic2022CD}. Additionally, 3.) acknowledging the critical role of temporal continuity in animation, we employ a roughness penalty regularization strategy aimed at ensuring smoother transitions between frames.

\section{Refined Inverse Rigging Methodology for Blendshape Animation}

In this section, we detail our approach for solving the inverse rig problem, prioritizing high accuracy in mesh reconstruction and ensuring smooth transitions between blendshape weights across frames. Unlike \cite{rackovic2022CD}, which focuses on single-frame analysis, our method evaluates the entire animation sequence, necessitating a matrix-based representation for weights and other elements in the objective function. 

\subsection{Data Fidelity and Regularization Framework}

Consider an animation comprising $T$ frames, denoted as $t=1, \ldots, T$. We represent the weight vectors for each frame as $\textbf{w}^{(t)}$ and assemble these into a weight matrix $\textbf{W} = [\textbf{w}^{(1)}, \ldots, \textbf{w}^{(T)}] \in \mathbb{R}^{m \times T}$. The rig function for the entire sequence is then expressed as:
%
%
\begin{equation}
\begin{split}
    f(\textbf{W}) = & \textbf{BW} +\sum_{t=1}^T \sum_{\{i,j\}\in\mathcal{P}}w_i^{(t)}w_j^{(t)}\text{diag}(\textbf{b}^{\{ij\}})\textbf{E}^{(t)}  + \sum_{\{i,j,k\}\in\mathcal{T}}w_i^{(t)}w_j^{(t)}w_k^{(t)}\text{diag}(\textbf{b}^{\{ijk\}})\textbf{E}^{(t)} \\ 
    & + \sum_{\{i,j,k,l\}\in\mathcal{Q}}w_i^{(t)}w_j^{(t)}w_k^{(t)}w_l^{(t)}\text{diag}(\textbf{b}^{\{ijkl\}})\textbf{E}^{(t)},
\end{split}
\end{equation}
where $\textbf{E}^{(t)}$ is a matrix of zeros with the $t^{th}$ column consisting of ones. Similarly, we define the target meshes matrix\\ $\hat{\textbf{B}} = [\hat{\textbf{b}}^{(1)}, \ldots, \hat{\textbf{b}}^{(T)}] \in \mathbb{R}^{n \times T}$, with $\hat{\textbf{b}}^{(t)}$ being the target mesh for frame $t$.
%
We propose the objective that consists of three terms:
\begin{equation}
    \minimize_{0\leq\textbf{W}\leq 1} E_{\text{df}} + \alpha E_{\text{sr}} + \beta E_{\text{tsr}},
\end{equation}
where $E_{\text{df}}$ stands for a data fidelity term, i.e., a difference between the estimated mesh and the target mesh in vertex space, $E_{\text{sr}}$ stands for the sparsity regularization forcing the cardinality of the weights to be low, and $E_{\text{tsr}}$ is temporal smoothness regularizer, forcing the weights in the consecutive frames to have similar values; $\alpha,\beta\geq 0$ are corresponding regularization weights, dictating the importance of each term. Let us observe each of these terms individually.

\paragraph{Data Fidelity}

The data fidelity term, along with the sparsity regularization term, aligns with the formulation in (\ref{eq:holistic_objective}) but is now adapted to a matrix context to handle the entire animation sequence. Specifically, the data fidelity term is defined as:
\begin{equation}\label{eq:data_fid_term}
    E_{\text{df}} = \frac{1}{n} \|f(\textbf{W}) - \hat{\textbf{B}}\|_F^2,
\end{equation}
where $\|\cdot\|_F$ denotes the Frobenius norm, measuring the discrepancy between the estimated mesh sequence $f(\textbf{W})$ and the target mesh sequence $\hat{\textbf{B}}$.

We employ a coordinate descent approach to minimize this term, focusing on a single blendshape controller, denoted as $e$, across the temporal dimension. In this context, we reformulate \eqref{eq:data_fid_term} into a quadratic expression:
\begin{equation*}
    E_{\text{df}} = \frac{1}{n} \textbf{W}_e^T \Phi \textbf{W}_e + \frac{2}{n} \textbf{W}_e^T \Theta,
\end{equation*}
where $\textbf{W}_e$ represents the weights corresponding to controller $e$ over all frames. The matrices $\Phi$ and $\Theta$ are constructed as follows:

Matrix $\Phi = \text{diag}([\phi^{(1)T}\phi^{(1)}, \ldots, \phi^{(T)T}\phi^{(T)}])$ encapsulates all the terms interacting with the quadratic terms of the objective. 
Each $\phi^{(t)}$ represents the contribution of the quadratic term of the blendshape controller $e$ at frame $t$, computed as:
\begin{equation}
        \phi^{(t)}_i = B_{ie} + \sum_{j \in \mathcal{P}(e)} w_j^{(t)} \textbf{b}^{\{je\}}_i + \sum_{\{j,k\} \in \mathcal{T}(e)} w_j^{(t)} w_k^{(t)} \textbf{b}^{\{jke\}}_i 
        + \sum_{\{j,k,l\} \in \mathcal{Q}(e)} w_j^{(t)} w_k^{(t)} w_l^{(t)} \textbf{b}^{\{jkle\}}_i,
\end{equation}
which considers not only the direct influence of controller $e$ but also its interaction with other controllers in the corrective terms.

Matrix $\Theta = [\phi^{(1)T}\psi^{(1)}, \ldots, \phi^{(T)T}\psi^{(T)}]^T$ represents the linear interaction terms, where each $\psi^{(t)}$ is given by:
\begin{equation}
    \begin{split}
        \psi_i^{(t)} &= \sum_{j \neq e} w_j^{(t)} B_{ij} + \sum_{\{j,k\} \in \mathcal{P}} w_j^{(t)} w_k^{(t)} \textbf{b}^{\{jk\}}_i + \sum_{\{j,k,l\} \in \mathcal{T}} w_j^{(t)} w_k^{(t)} w_l^{(t)} \textbf{b}^{\{jkl\}}_i \\
        &+ \sum_{\{j,k,l,h\} \in \mathcal{Q}} w_j^{(t)} w_k^{(t)} w_l^{(t)} w_h^{(t)} \textbf{b}^{\{jklh\}}_i - \hat{\textbf{b}}^{(t)}_i.
    \end{split}
\end{equation}
This accounts for the contributions of all other blendshape controllers, as well as the deviation from the target mesh $\hat{\textbf{b}}^{(t)}$ at frame~$t$.

Through this formulation, the data fidelity term effectively quantifies and minimizes the difference between the animated mesh sequence and the target sequence, thereby ensuring high accuracy in replicating desired facial expressions over time.

\paragraph{Sparsity Regularization}

In our approach, the sparsity regularization term is critical for ensuring that the animation remains computationally efficient and interpretable. It is defined as the normalized sum of all blendshape weights across the entire animation sequence. Mathematically, this is represented as:
\begin{equation*}
    E_{\text{sr}} = \frac{1}{m} \sum_{i=1}^m \sum_{t=1}^T w_i^{(t)} = \frac{1}{m} \textbf{1}^T \textbf{W} \textbf{1},
\end{equation*}
where $\textbf{1}$ denotes a vector of ones of appropriate dimension. This formulation encourages the model to use as few active blendshapes as possible at each frame, leading to a sparser and more interpretable set of blendshape weights. Importantly, due to the non-negativity constraints on the weights, $E_{\text{sr}}$ is guaranteed to be a non-negative term.

The integration of sparsity regularization into the optimization process serves multiple purposes: it not only enhances computational efficiency by reducing the number of active blendshapes but also simplifies the task of manual adjustments or further processing by animators. By penalizing the sum of the weights, the model naturally gravitates towards solutions where fewer blendshapes are used to achieve the desired facial expressions, thereby promoting a more streamlined and manageable animation process.

\paragraph{Temporal Smoothness Regularization}

A key aspect of realistic animation is the smoothness of transitions between frames. To achieve this, we incorporate a roughness penalty function into our regularization framework. This function penalizes the squared differences between adjacent weight values across frames, effectively encouraging temporal continuity in the animation. The temporal smoothness regularization term is formulated as follows:
\begin{equation}
    E_{\text{tsr}} = \sum_{t=1}^{T-2} \sum_{i=1}^m |w_i^{(t)} - 2w_i^{(t+1)} + w_i^{(t+2)}|^2 = \sum_i \textbf{W}_i^T \textbf{F} \textbf{W}_i,
\end{equation}
where $\textbf{W}_i$ denotes the weight vector for the $i^{th}$ blendshape across all frames. The matrix $\textbf{F}$ is a pentadiagonal matrix defined as:
\begin{equation}
    \textbf{F}=
    \begin{bmatrix}
        1 & -2 & 1 & \color{gray} 0 & \cdots & \color{gray} 0 \\
        -2 & 5 & -4 & 1 & \cdots & \color{gray} 0 \\
        1 & -4 & 6 & -4 & \ddots & \vdots \\
        \color{gray} 0 & 1 & -4 & 6 & \ddots & 1 \\
        \vdots & \ddots & \ddots & \ddots & \ddots & -2 \\
        \color{gray} 0 & \cdots & \color{gray} 0 & 1 & -2 & 1
    \end{bmatrix},
\end{equation}
with the pattern designed to penalize the roughness or abrupt changes in the weight vectors between consecutive frames.

This regularization term thus plays a crucial role in ensuring the naturalness and fluidity of the generated animation. By minimizing $E_{\text{tsr}}$, our model actively works to smooth out the transitions, leading to more lifelike and appealing animations that closely mimic natural human expressions over time.

\paragraph{Final formulation.}
To synthesize the various aspects of our method into a coherent optimization framework, we formulate a comprehensive objective function that balances the need for data fidelity, sparsity, and temporal smoothness. This objective, to be minimized with respect to each blendshape controller $e$, encapsulates the essence of our approach:
\begin{equation}\label{eq:derived_objective}
    \minimize_{\textbf{0} \leq \textbf{W}_e \leq \textbf{1}} \textbf{W}_e^T \left(\frac{1}{n} \Phi + \beta \textbf{F} \right) \textbf{W}_e + 2 \textbf{W}_e^T \left(\frac{1}{n} \Theta + \frac{\alpha}{2m} \textbf{1} \right).
\end{equation}

In this equation, $\Phi$ and $\Theta$ are matrices derived from the data fidelity term, encoding the relationship between the blendshape weights and the target meshes. The matrix $\textbf{F}$, arising from the temporal smoothness regularization term, ensures that changes in the blendshape weights are gradual over the sequence of frames. Lastly, the term involving $\textbf{1}$, originating from the sparsity regularization, promotes solutions with fewer active blendshapes, thereby aiding in interpretability and computational efficiency.

This carefully constructed objective function is central to our method, guiding the optimization process towards solutions that are not only accurate in reproducing the target facial expressions but also efficient and smooth over time. By balancing these critical aspects, our approach advances the state-of-the-art in blendshape animation, particularly in scenarios requiring high fidelity and natural motion dynamics.

\subsection{Clustering Approach for Computational Efficiency}
\label{sec:clustering}
\begin{figure}
    \centering
    \includegraphics[width=0.66\linewidth]{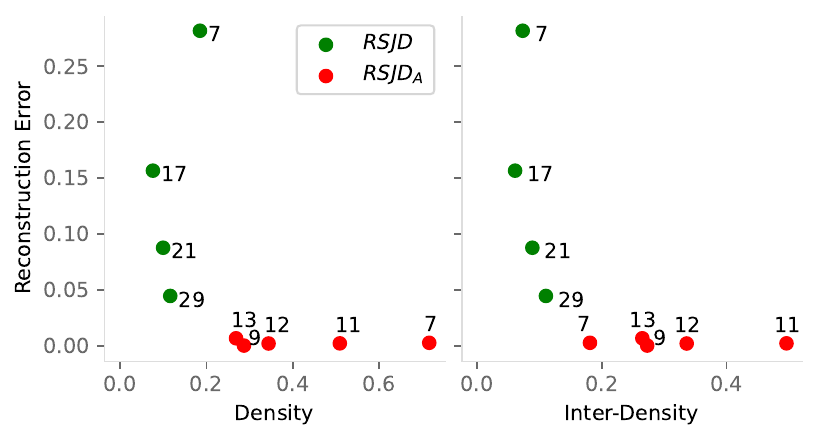}
    \includegraphics[width=0.66\linewidth]{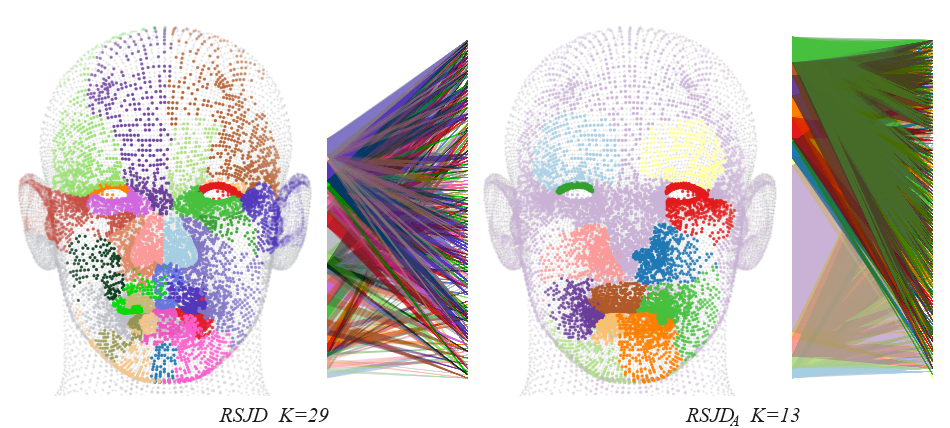}
    \caption{\textbf{Top row}: The trade-off between \textit{Reconstruction Error} ($E_R$) and \textit{Density} ($E_D$) (left), and \textit{Inter-Density} ($E_{ID}$) (right), across different clustering approaches, with annotations indicating the chosen number of clusters ($K$). \textbf{Bottom row}: Visualization of clusters obtained using $RSJD$ with $K=29$ (left) and $RSJD_A$ with $K=13$ (right). In addition to the mesh clusters, a bipartite graph representation is shown, using the same color coding, where the left partition denotes mesh vertices, and the right partition signifies the blendshape indices assigned to each cluster.}
    \label{fig:clusters}
\end{figure}
Given the localized nature of facial features, a significant number of vertices in the human face may have limited influence on the majority of blendshape weights. Our approach incorporates a face clustering strategy to exploit this characteristic for enhanced computational efficiency. This strategy allows for a parallelized and potentially more regularized solution to the inverse rigging problem \cite{seol2011artist, rackovic2021clustering, rackovic2023distributed}.

We adopt the clustering methodologies proposed in \cite{rackovic2021clustering} (denoted as $RSJD$) and \cite{rackovic2023distributed} (denoted as $RSJD_A$). These methods are particularly suitable as they generate clusters in a model-based manner. However, our approach remains flexible enough to be applicable with other clustering techniques, provided they meaningfully separate mesh segments and blendshape controllers.

The core idea is to treat each cluster as an independent model, thereby decomposing the overall inverse rigging problem (\ref{eq:derived_objective}) into a series of smaller subproblems. Each subproblem focuses on a specific subset of vertices and blendshapes relevant to its cluster. This subdivision not only facilitates parallel processing but also potentially introduces additional regularization opportunities, leading to a more efficient solution overall.

Following the recommendations in \cite{rackovic2023distributed}, we generate several instantiations of each clustering approach. The optimal configuration is selected based on a balance between \textit{Reconstruction Error} and \textit{Density} in the resulting clustering graphs, as illustrated in Figure \ref{fig:clusters} (top). Optimal cluster numbers, such as $K=29$ for $RSJD$ and $K=13$ for $RSJD_A$, are determined to minimize the model size while retaining essential information. The corresponding mesh clusters and their bipartite graph representations are detailed in Figure \ref{fig:clusters} (bottom), showcasing the effective distribution of vertices and blendshape indices across the identified clusters.

\subsection{Optimization Strategy}

The optimization strategy used in solving the proposed objective is \textit{coordinate descent}, a well established optimization technique that is guaranteed to produce monotonically non-increasing costs \cite{luo1992convergence, wright2015coordinate}. With coordinate descent, a single component of the problem is visited at a time, and the objective is minimized in it, while keeping the other values fixed, as we do in \eqref{eq:derived_objective}, observing only a single controller $e$ as a variable. This decoupling in components allows us to simplify otherwise non-convex problem \eqref{eq:our_objective}, into a constrained quadratic program, for which fast solutions and implementations are readily available \cite{more1989algorithms, virtanen2020scipy}. Coordinate descent is particularly suitable for for the application in inverse rigging, since estimating one blendshape weight at a time will help avoid the simultaneous activation of blendshapes with canceling effects, i.e., moving the corner of the mouth up and down at the same time \cite{seol2011artist}. 

An important aspect of coordinate descent is the order of component updates. Even though the convergence guaranties hold with an arbitrary update order \cite{wright2015coordinate}, in practice a poor choice can lead to slower convergence and a bad local minima. This problem was also studied in blendshape animation literature \cite{seol2011artist,rackovic2022CD,hyde2021obtaining}. We follow the stance of \cite{seol2011artist}, ordering the blendshapes by their overall magnitude of deformation in the mesh, in line with artist intuition of initially setting the more drastic weights before focusing on fine details. 



\section{Quantitative Analysis of Animation Techniques}
This section presents the comprehensive evaluation of our proposed blendshape animation methodology. We first describe the data and benchmark models used in our study, followed by a detailed analysis of the results. The performance of our method is compared against established benchmarks to demonstrate its efficacy in achieving high-fidelity, smooth animations. The impact of different parameters on the results is also examined to provide insights into the behavior of our approach under various conditions.

\begin{figure}
    \centering
    \includegraphics[width=0.66\linewidth]{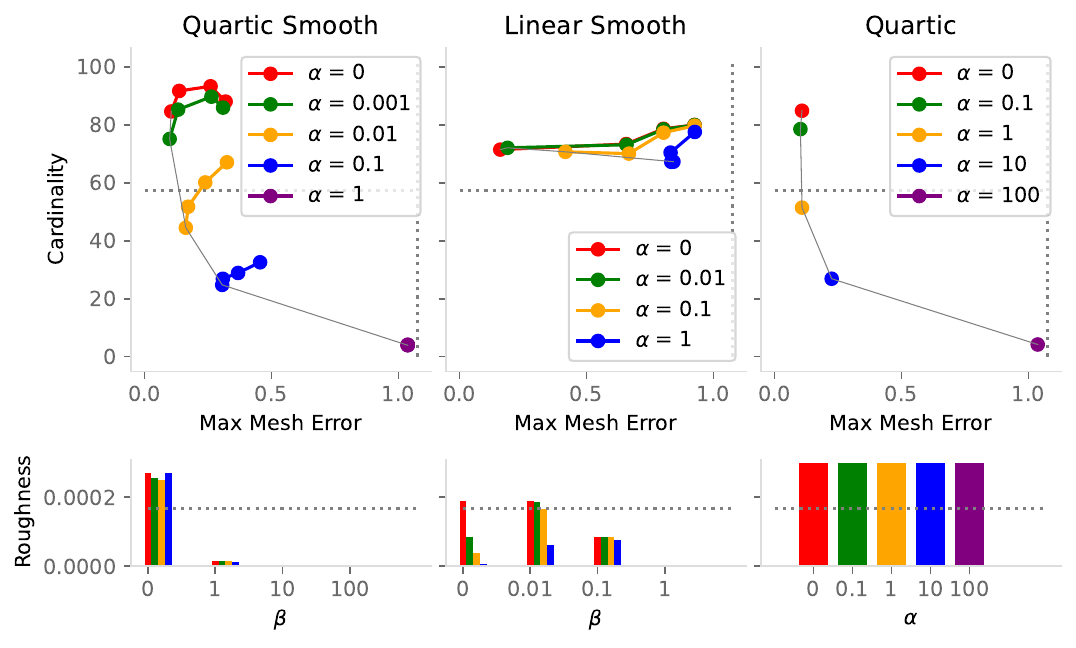}
    \caption{Parametric Analysis of Animation Metrics During Training, comparing our method (\textit{Quartic Smooth}) with benchmarks. \textbf{Top row:} This graph illustrates the interplay between blendshape cardinality and maximum mesh error under various animation approaches. The color coding denotes different levels of the sparsity regularizer, $\alpha$, while individual points represent a spectrum of the smoothness parameter, $\beta$, converging at $\beta=0$ indicated by the solid gray line. The horizontal dotted line marks the cardinality of the actual animation data used as the ground truth, whereas the vertical dotted line indicates the mesh error that would result if no blendshapes were activated (all weights at $0$). \textbf{Bottom row:} Here, we chart the roughness penalty corresponding to the varying values of $\beta$ along the x-axis. The color scheme is consistent with the top graph, linked to the $\alpha$ values. The horizontal dotted line represents the benchmark roughness penalty derived from the ground-truth animation data.}
    \label{fig:holistic_training}
\end{figure}


\begin{figure}
    \centering
    \includegraphics[width=0.66\linewidth]{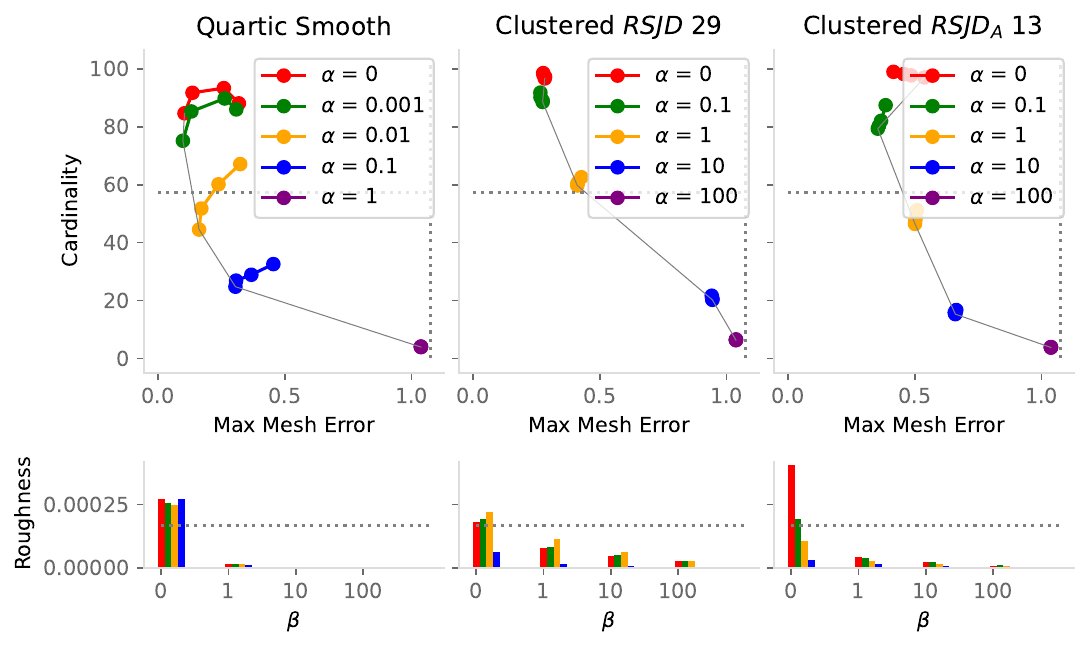}
    \caption{Comparative Analysis of Training Metrics Across Different Parameterizations and Methodologies. In this Figure we show how applying  clustering on top of our approach affects the overall results.}
    \label{fig:clustered_training}
\end{figure}


\subsection{Dataset Characteristics and Comparative Benchmarks}

We selected the Metahuman character \textit{Jesse} to evaluate our algorithm. This choice is motivated by the high-quality and realistic nature of Metahuman blendshape models, coupled with their accessibility as publicly available resources. The \textit{Jesse} model is equipped with $m=80$ base blendshapes and over $400$ corrective terms, encompassing the first, second, and third levels of correction. These elements make it an ideal candidate for demonstrating the nuanced capabilities of our method in handling complex facial animations.

Originally, the meshes contain $24000$ vertices. To focus on the facial region, which is our primary interest, we have excluded vertices on the neck and an inactive area at the back of the skull. This results in a subsampled mesh comprising $n=10000$ vertices. The model is animated to generate a reference motion, which comprises $80$ frames for the training set and $100$ frames for the test set. This division allows for a robust assessment of our method's performance both in learning and generalization.

In our results section, we refer to our proposed method as \textit{Quartic Smooth}. This designation emphasizes its unique aspects: the inclusion of corrective terms using a quartic blendshape function for enhanced realism, and the integration of smoothness regularization, setting it apart from previous model-based solutions. These characteristics are pivotal in achieving more lifelike and dynamically consistent facial animations.

Our benchmarks include two prior works for comparative analysis. The first is the method described in \cite{seo2011compression}, named \textit{Linear Smooth}, which incorporates smoothness regularization but does not include corrective terms. This method optimizes the objective defined in (\ref{eq:seo_objective}). The second benchmark is the \textit{Quartic} algorithm from \cite{rackovic2023distributed}, which considers corrective terms for better mesh reconstruction but treats each frame independently, overlooking temporal continuity. \textit{Quartic} solves the objective in (\ref{eq:holistic_objective}). Notably, while both benchmarks include weight regularization, they differ in approach: \textit{Quartic} uses an $l_1$ norm, promoting sparsity, whereas \textit{Linear Smooth} employs a squared $l_2$ norm, penalizing large activations without directly reducing the number of active components.

Additionally, we explore an extension of our proposed algorithm that incorporates the face clustering technique detailed in Section \ref{sec:clustering}. We examine two instances of clustering: \textit{Clustered $RSJD$ 29} and \textit{Clustered $RSJD_A$ 13}, named after the clustering methods and the chosen number of clusters ($K$). These cases help to assess the impact of facial clustering on the performance and efficiency of our approach.

\subsection{Analytical Performance Evaluation}

All the considered methods have a weight regularization hyperparameter $\alpha$ that should be selected before comparing the results in more details. Additionally, our proposed method, \textit{Quartic Smooth}, as well as the benchmark \textit{Linear Smooth}, have the smoothness regularization hyperparameter $\beta$. To select the appropriate values, each method is cross-validated with a wide choice of values, as presented in Fig. \ref{fig:holistic_training} and \ref{fig:clustered_training}. In general these results draw an expected trade-of curve between cardinality and mesh error, with the exception of the \textit{Linear Smooth} --- in this case a favorable decrease of the cardinality is never achieved, since $l_2$ norm in the objective tends to keep small positive values rather than setting them to zero. Optimal values for the parameters are selected so that all three presented metrics are minimized, and the final selection that is used for further tests is listed in Tab. \ref{tab:test_parameters}.

\begin{table}[h]
\centering
\begin{tabular}{l|c c c c c}
          & \makecell{Quartic \\ Smooth}  & \makecell{Linear \\ Smooth} & \makecell{Quartic}  & \makecell{Clustered \\ $RSJD$ 29} & \makecell{Clustered \\ $RSJD_A$ 13}\\ \hline
$\alpha$  & $0.0078$ &  $0.01$    & $0.9$ &  $1$  &  $0.8$  \\ 
$\beta$   & $1$      &  $0$       &  /    &  $10$ &  $1$    \\ 
\end{tabular}
\caption{Final parameter values for each method.}
\label{tab:test_parameters}
\end{table}

\begin{figure}
    \centering
    \includegraphics[width=0.66\linewidth]{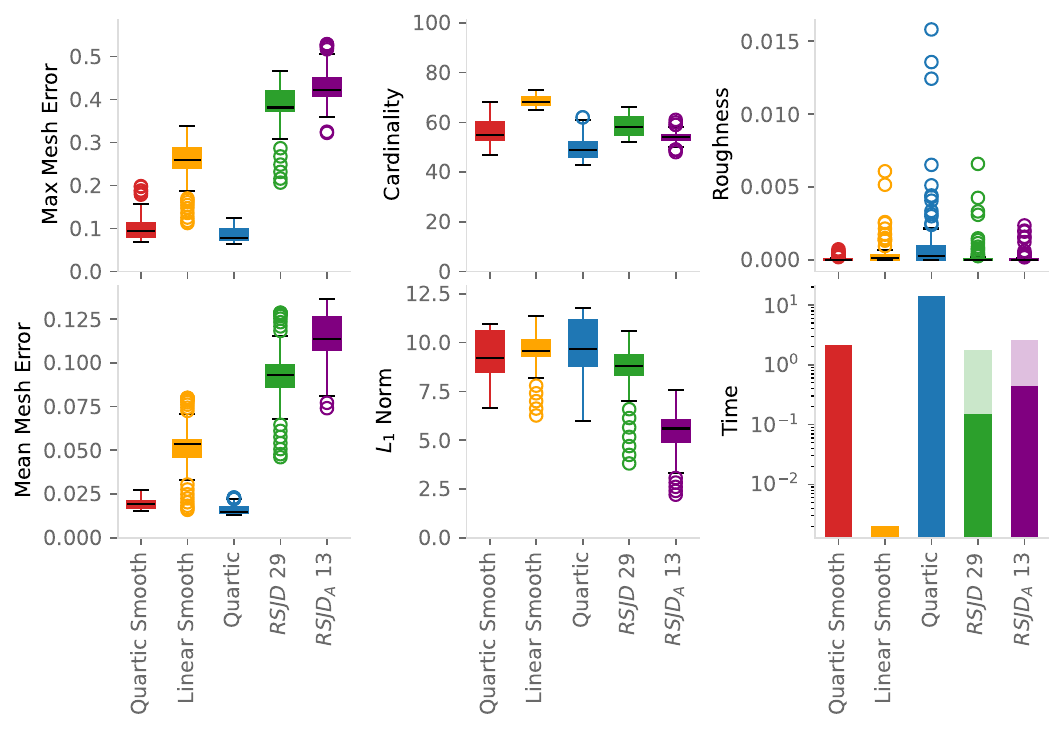}
    \caption{Results over the test set with the selected hyperparameter values corresponding to Table \ref{tab:test_parameters}. The execution time for the clustered approach is presented in solid and shaded --- solid color indicates the execution time of the slowest cluster, as that is the cost when solving the problem in parallel, while shaded bar shows the time of solving the clusters sequentially.}
    \label{fig:test_res}
\end{figure}

\begin{table}[h]
\centering
\begin{tabular}{r|c c c c c c}
    & {\makecell{Max Mesh \\ Error}} & {\makecell{Mean Mesh \\ Error}} & {Cardinality} & {$l_1$ Norm} & {\makecell{Roughness \\ Penalty}} & {\makecell{Execution \\ Time}} \\ \hline
    \makecell{Quartic \\ Smooth}  & $.101$ & $.019$ & $56.3$ & $9.431$ & $7.2e^{-5}$ & $2.16$ \\ 
    \makecell{Linear \\ Smooth}   & $.257$ & $.051$ & $68.6$ & $9.649$ & $4.2e^{-4}$ & $0.02$ \\ 
    Quartic                       & $.086$ & $.016$ & $49.7$ & $9.818$ & $1.1e^{-3}$ & $14.4$ \\ 
    $RSJD$                        & $.387$ & $.093$ & $58.2$ & $8.695$ & $2.8e^{-4}$ & $0.15$ \\ 
    $RSJD_A$                      & $.433$ & $.115$ & $54.0$ & $5.500$ & $1.5e^{-4}$ & $0.44$ \\ 
\end{tabular}
\caption{Average metrics values for test results (see also Figure \ref{fig:test_res}).}
\label{tab:test_numerical_res}
\end{table}
By examining Tab.~\ref{tab:test_numerical_res}, we can compare the performance of the various methods (\textit{Quartic Smooth}, \textit{Linear Smooth}, \textit{Quartic}, \textit{RSJD}, and \textit{$RSJD_A$}) across several key metrics: Max Mesh Error, Mean Mesh Error, Cardinality, $l_1$ Norm, Roughness Penalty, and Execution Time.
\paragraph*{Max Mesh Error and Mean Mesh Error:} \textit{Quartic} exhibits the lowest Max and Mean Mesh Errors ($.086$ and $.016$, respectively), indicating its superior ability to closely match the target mesh in the worst-case and on average. This suggests high accuracy in mesh reconstruction. 
\textit{Quartic Smooth} also performs well, with slightly higher errors, but still significantly better than \textit{Linear Smooth}, \textit{RSJD}, and \textit{$RSJD_A$}.
\paragraph*{Cardinality:} The Cardinality, which indicates the number of active blendshapes, is lowest for \textit{Quartic} (49.7), suggesting it is the most efficient in terms of blendshape usage. \textit{Quartic Smooth} has a moderately higher cardinality (56.3) compared to \textit{Quartic}, but lower than \textit{Linear Smooth} and \textit{RSJD}.
\paragraph*{$l_1$ Norm:} The $l_1$ Norm, reflecting the sum of absolute weights, is highest for \textit{Quartic} (9.818) and lowest for \textit{$RSJD_A$} (5.500). Higher values indicate more weight is being used overall, which can suggest more intense or complex facial expressions.
\paragraph*{Roughness Penalty:} \textit{Quartic Smooth} has the lowest Roughness Penalty ($7.2e^{-5}$), highlighting its effectiveness in ensuring smooth frame-to-frame transitions. \textit{Quartic}, despite its accuracy, has a higher roughness penalty ($1.1e^{-3}$), implying less smoothness in transitions compared to \textit{Quartic Smooth}.
\paragraph*{Execution Time:} \textit{Linear Smooth} is the fastest ($.02$ seconds), which is expected given its less complex nature (lacking corrective terms). \textit{Quartic} is the slowest (14.4 seconds), possibly due to the computational overhead of handling corrective terms. \textit{Quartic Smooth} strikes a balance between complexity and speed (2.16 seconds), offering a more efficient solution than \textit{Quartic} while maintaining high accuracy and smoothness. This advantage over \textit{Quartic} is due to the parallelizable structure given by the clustering.

\textit{Quartic Smooth} demonstrates a well-balanced performance across accuracy, efficiency, and smoothness, making it a robust choice for high-quality and realistic blendshape animation. \textit{Quartic} excels in mesh error metrics but at the cost of higher roughness and longer execution time, while \textit{Linear Smooth} offers the fastest execution with compromises in accuracy and smoothness. The clustered approaches (\textit{RSJD} and \textit{$RSJD_A$}) offer varying trade-offs, with \textit{$RSJD_A$} showing lower execution times and roughness penalties but at the cost of higher mesh errors.

Figure~\ref{fig:holistic_training} shows graphs evaluating three approaches: Quartic Smooth, Linear Smooth, and Quartic. Each graph shows the trade-off between cardinality and maximum mesh error on the top row and the roughness penalty across different smoothness parameter $\beta$ values on the bottom row. 
\paragraph*{Quartic Smooth:} The top graph suggests that as the sparsity regularizer 
$\alpha$ increases, both the cardinality and the maximum mesh error generally increase. This indicates that higher sparsity (achieved by a larger $\alpha$) comes at the cost of accuracy (higher mesh error).
In the bottom graph, the roughness penalty seems relatively stable across various values of $\beta$, suggesting that the smoothness regularization is not significantly impacting the roughness penalty in this approach.
\paragraph*{Linear Smooth:} The top graph shows that the maximum mesh error and cardinality do not significantly change with different values of $\alpha$, indicating a potential plateau in the trade-off, where increasing sparsity does not impact the mesh error substantially.
The bottom graph shows variations in the roughness penalty across different values of $\beta$, but the changes are not dramatic, implying that the model's smoothness is not highly sensitive to this parameter within the tested range.
\paragraph*{Quartic:} The top graph indicates that increasing 
$\alpha$ leads to a decrease in the maximum mesh error but at the expense of a rapid increase in cardinality, suggesting a strong influence of the sparsity regularizer on model complexity.
There's no bottom graph for roughness as the Quartic model does not seem to include the smoothness parameter $\beta$. 

The Quartic Smooth approach seems to strike a balance between maintaining mesh accuracy and controlling the model complexity (cardinality), especially when compared to the Linear Smooth and Quartic approaches. The roughness penalties across all methods are relatively low, indicating smooth transitions, with Quartic Smooth showing a slight advantage. However, one must consider the trade-offs between accuracy, sparsity, and temporal smoothness when selecting the appropriate parameters for each method.

Figure~\ref{fig:clustered_training} shows a comparative analysis of three different blendshape animation approaches: Quartic Smooth, Clustered RSJD 29, and Clustered RSJD\_A 13. The analysis is based on the evolution of specific metrics over the training set while varying two parameters: the sparsity regularizer ($\alpha$) and the smoothness parameter ($\beta$).

The Quartic Smooth approach in our experiments offers a favorable balance between animation accuracy and sparsity, with smooth temporal transitions as indicated by the low roughness penalties.
The clustered approaches may provide computational benefits, but potentially at the cost of increased max mesh error, especially for higher sparsity levels.
The choice between these methods may depend on the specific requirements of an animation project, such as the necessity for real-time performance (favoring clustered methods) versus the need for high-fidelity animations (favoring Quartic Smooth).

\begin{figure*}
    \centering
    \includegraphics[width=0.9\linewidth]{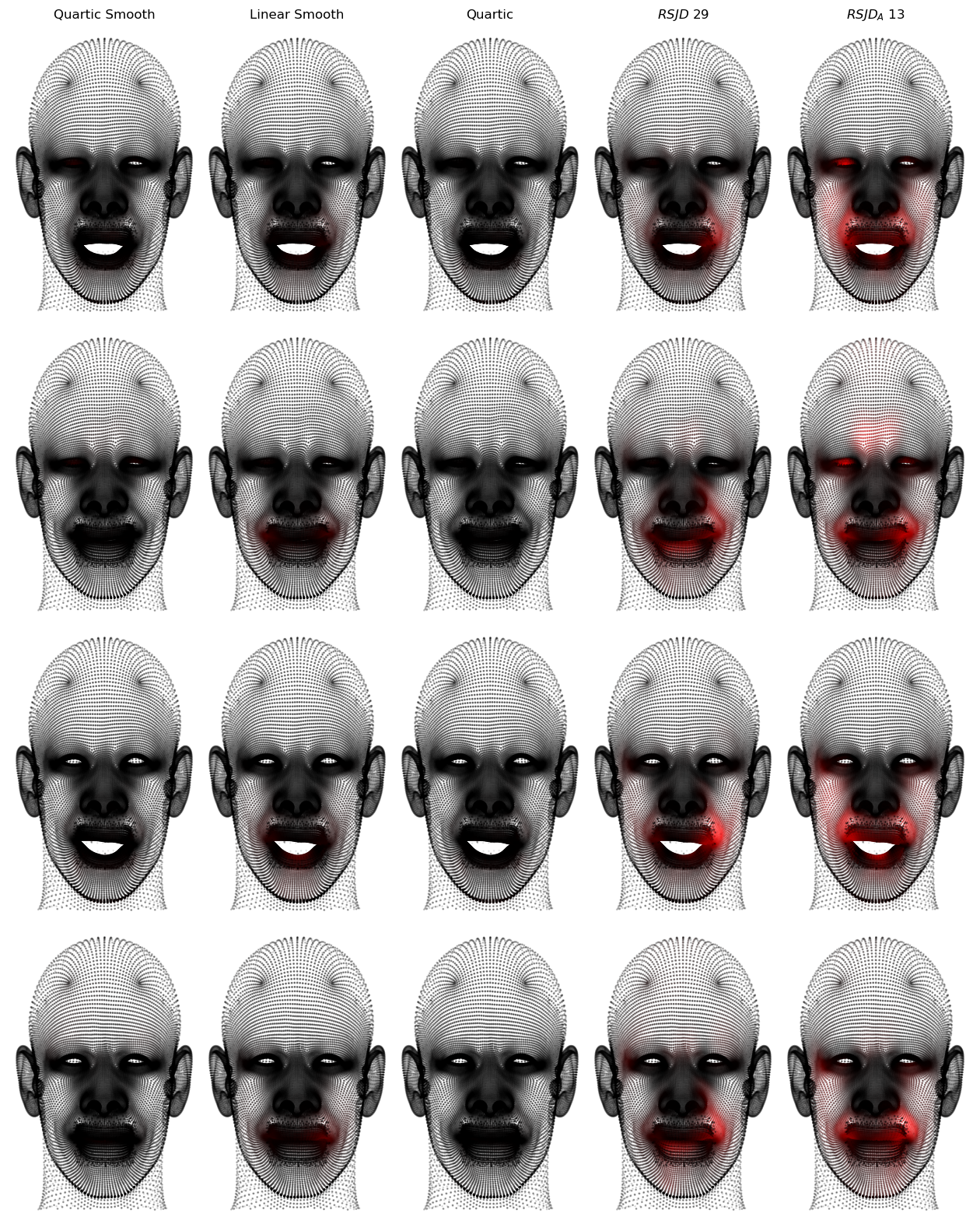}
    \caption{Resulting meshes for selected frames (rows), and for each method (columns). Stronger red tones indicate higher mesh error.}
    \label{fig:error_scatterplots}
\end{figure*}

\section{Conclusion}

We have introduced \textit{Quartic Smooth}, an advanced model-based approach for blendshape animation that adeptly balances accuracy, sparsity, and temporal smoothness. Our method demonstrates a marked improvement in mesh fidelity over traditional linear models and offers a flexible framework for both high-quality and real-time applications.
Our findings illustrate \textit{Quartic Smooth}'s superior performance in creating realistic facial animations with reduced computational overhead, particularly when compared to existing linear and non-linear approaches. The introduction of face clustering techniques further augments computational efficiency, opening possibilities for real-time animation processing.
Future work will explore the optimization of these techniques and the integration of machine learning to refine parameter selection. With its contribution to the realistic and efficient inverse rigging of facial animations, \textit{Quartic Smooth} is poised to influence future developments in character animation within the computer graphics community.

\bibliographystyle{unsrtnat}
\bibliography{references} 

\end{document}